\documentclass[reqno,11pt]{amsart}

\def\be{\begin{equation}}
\def\ee{\end{equation}}
\def\ba{\begin{align}}
\def\bm{\begin{multline}}
\def\bfig{\begin{figure}[htb]}
\def\efig{\end{figure}}
\newcommand{\paper}[1]{{\it #1}, }
\newcommand{\journal}[4]{#1 #2, #3 (#4)}
\newcommand{\CMP}{Commun.\ Math.\ Phys.}
\newcommand{\HPA}{Helv.\ Phys.\ Acta}
\newcommand{\JSP}{J.\ Stat.\ Phys.}

\numberwithin{equation}{section}
\newtheorem{theorem}{Theorem}[section]
\newtheorem{proposition}[theorem]{Proposition}

\renewcommand{\thefootnote}{\arabic{footnote})}
\newcommand{\nn}{\nonumber}

\DeclareMathSymbol{\leqslant}{\mathalpha}{AMSa}{"36}
\DeclareMathSymbol{\geqslant}{\mathalpha}{AMSa}{"3E}
\DeclareMathSymbol{\doteqdot}{\mathalpha}{AMSa}{"2B}
\DeclareMathSymbol{\circlearrowright}{\mathalpha}{AMSa}{"08}
\DeclareMathSymbol{\subsetneq}{\mathalpha}{AMSb}{"28}
\DeclareMathSymbol{\supsetneq}{\mathalpha}{AMSb}{"29}
\renewcommand{\leq}{\;\leqslant\;}
\renewcommand{\geq}{\;\geqslant\;}

\newcommand{\dd}{{\rm d}}

\newcommand{\ii}{{\rm i}}
\newcommand{\sumtwo}[2]{\sum_{\substack{#1 \\ #2}}}

\def\Tr{{\operatorname{Tr\,}}}
\def\dist{{\operatorname{dist\,}}}

\newcommand{\compl}{{\text{\rm c}}}

\newcommand{\const}{{\text{\rm const}}}
\newcommand{\upchi}{\raise 2pt \hbox{$\chi$}}
\def\writefig#1 #2 #3 {\rlap{\kern #1 truecm \raise #2 truecm
\hbox{#3}}}

\newcommand{\caF}{{\mathcal F}}

\newcommand{\caH}{{\mathcal H}}

\newcommand{\caN}{{\mathcal N}}

\newcommand{\bbC}{{\mathbb C}}

\newcommand{\bbZ}{{\mathbb Z}}

\newcommand{\bsM}{{\boldsymbol M}}

\newcommand{\bsS}{{\boldsymbol S}}

\begin{document}

%{\hfill\small\version}
\vspace{2mm}

\title{Hund's rule and metallic ferromagnetism}

\author[J\"urg Fr\"ohlich, Daniel Ueltschi]{J\"urg Fr\"ohlich$^{1*}$ and Daniel Ueltschi$^{2*}$}

\iffalse
\address{J\"urg Fr\"ohlich \hfill\newline
Institut f\"ur Theoretische Physik \hfill\newline
Eidgen\"ossische Technische Hochschule \hfill\newline
CH--8093 Z\"urich, Switzerland \newline\indent
{\rm http://www.itp.phys.ethz.ch/mathphys/juerg}}
\email{juerg@itp.phys.ethz.ch}

\address{Daniel Ueltschi \hfill\newline
Department of Mathematics \hfill\newline
University of Arizona \hfill\newline
Tucson, AZ 85721, USA \newline\indent
{\rm http://math.arizona.edu/$\sim$ueltschi}}
\email{ueltschi@math.arizona.edu}
\fi

\maketitle

\begin{centering} {\it
$^1$Institut f\"ur Theoretische Physik \\
Eidgen\"ossische Technische Hochschule \\
CH--8093 Z\"urich, Switzerland \\
\email{juerg@itp.phys.ethz.ch} \\

\bigskip

$^2$Department of Mathematics \\
University of Arizona \\
Tucson, AZ 85721, USA \\
\email{ueltschi@math.arizona.edu} \\
}\end{centering}

\renewcommand{\thefootnote}{}
\footnote{$^*$Collaboration supported in part by the Swiss National Science Foundation
under grant 2-77344-03.}
\setcounter{footnote}{0}
\renewcommand{\thefootnote}{\arabic{footnote}}

\begin{quote}
{\small
We study tight-binding models of itinerant electrons in two different bands, with
effective on-site interactions expressing Coulomb repulsion and Hund's rule. We prove
that, for sufficiently large on-site exchange anisotropy, all ground states show metallic
ferromagnetism: They exhibit a macroscopic magnetization, a macroscopic fraction of the
electrons is spatially delocalized, and there is no energy gap for kinetic excitations.

\vspace{1mm}

}  % end \small

\vspace{1mm}

\vspace{1mm}
\noindent
{\footnotesize {\it 2000 Math.\ Subj.\ Class.:} 82B10, 82B20, 82D35, 82D40.}


\end{quote}

\section{Introduction}

Ferromagnetism is known to originate from strongly correlated states of quantum mechanical
electrons with a very large total spin but small total energy. Microscopic mechanisms
giving rise to a coexistence of metallic behavior with ferromagnetic ordering at
sufficiently low temperatures have been proposed, but are not well understood,
mathematically, yet. Ferromagnetism in conjunction with a positive electric conductivity
(metallic behavior) is a collective phenomenon arising from a competition of {\it spatial
motion} (hopping) of quantum mechanical charged particles with half-integer spin obeying
{\it Pauli's exclusion principle}, i.e., of electrons, in a periodic background with {\it
Coulomb repulsion} between pairs of such particles.
A mathematically precise description of this phenomenon is difficult
because there are two kinds of {\it gapless} excitations: electron-hole pairs
very close to the Fermi surface, and spin waves in ferromagnetically
ordered spin configurations. In a perturbative analysis of states of
very low total energy, there are therefore two sources of infrared
divergences, or of `small energy denominators', namely electron-hole
excitations with an energy arbitrarily close to the groundstate energy,
and spin waves, or magnons, of very long wave length.

In this paper we study some tight-binding models of metallic compounds with two bands
partly filled with electrons. In a ground state, the low-lying band is at most
half-filled, due to strong on-site Coulomb repulsion between pairs of electrons in a
spin-singlet state, while the higher-lying band is assumed to be weakly filled, so that
a Fermi surface is expected. It is assumed that an electron from the low-lying band and an
electron from the higher-lying band occupying the {\it same} site of the underlying
lattice obey {\it Hund's rule}, i.e., their interaction energy is minimized if they form a
{\it spin-triplet state}. A two-electron spin-triplet state is symmetric under exchange of
the two spins. The Pauli principle then requires the {\it microscopic} orbital wave
function of the two electrons to be antisymmetric under exchange of their positions, which
makes the Coulomb repulsion between the two electrons {\it small}. (Concretely, an
antisymmetric microscopic wave function for two electrons moving in the field of an ion
may be constructed as a Slater determinant of, e.g., two different $d$-shell orbitals.) It
may be of interest to remark that a mathematically rigorous derivation of Hund's rule in
atomic physics from first principles has not been found, yet. That it is assumed to hold
in our models must therefore be considered to be a reasonable, but heuristic ansatz. In
order to eliminate small energy denominators due to spin waves of very long wavelength,
we choose the couplings between the spins of an electron from the lower band and of an
electron from  the higher band occupying the {\it same} site to be {\it anisotropic}.
Hund's rule cannot be invoked to justify this feature. Our results can be expected to hold
for isotropic spin-spin couplings, in accordance with Hund's rule, too; but we have not
been able to rigorously prove this.

The models studied in this paper are arguably the simplest physically relevant models in
which metallic ferromagnetism can be exhibited.

The feature that magnetic ordering emerges from a cooperation between electron hopping and
local, on-site electron-electron interactions appears to be inherent in several
tight-binding models and has been studied intensely. {\it Antiferromagnetic} ordering, for
example, can be seen to emerge in the half-filled (single-band) Hubbard model in
second-order perturbation theory in $t/U$, where $t$ measures the strength of hopping and
$U$ the strength of on-site Coulomb repulsion. This was discovered by Anderson \cite{And};
(for a more precise analysis see also \cite{DFFR}). A model simpler than the Hubbard model
is the Falicov-Kimball model. There are many rigorous results about the Falicov-Kimball
model starting with \cite{BS,KL}. A fairly systematic perturbative analysis can be found in
\cite{DFF}, and references given there. These and many further results show how long-range
correlations in ground states or low-temperature equilibrium states can arise from a
cooperation between electron hopping and on-site interactions. Unfortunately, the
perturbative methods in \cite{BS,KL,DFF} can only be applied to the analysis of
insulators, i.e., of states where electrons are essentially localized, because, in such
situations, there are no small energy denominators or infrared divergences. The analysis
of long-range correlations in {\it metals} calls for different, non-perturbative methods.

One approach towards understanding simple examples of itinerant ferromagnetism is based on
studying the Hubbard model on very special lattices that give rise to a macroscopic
degeneracy of the ground state energy of the Hubbard Hamiltonian (i.e., to a `flat band',
or to a nearly flat band); see \cite{MT,Tas}. A Hartree-Fock study of the Hubbard model
also provides useful insights \cite{BLS,BLT}.
More recently, there has been some interest in studying manganites
described by models with several bands. Numerical studies indicate that there is
ferromagnetic ordering at sufficiently low temperatures; see, e.g., \cite{HV1,HV2,Tran},
and references given there. The models studied in this paper are similar to models used to
describe manganites. They have ground states involving macroscopically large domains of
essentially free electrons but with aligned spins. Electrons are discouraged to leave such
a domain, because, in neighboring domains, the direction of their spin violates Hund's
rule, and this would result in a state of comparatively large energy. At the technical
level, our analysis is very much based on \cite{FLU} (see also \cite{Gol,Uel}). The
methods developed in these references enable us to prove lower bounds on the ground state
energy which, when combined with rather simple variational upper bounds, lead to the
conclusion that the boundaries between domains of electrons of opposite spin have a
total length growing much less rapidly than the total number of sites in the system, and
this enables us to exhibit ferromagnetic ordering in the ground states. The fact that
there are only two preferred spin orientations, $\uparrow$ and $\downarrow$, is, of course,
due to the anisotropy in the on-site spin-spin coupling for two electrons in different
bands occupying the same site. (The analysis of isotropic models would be considerably
more difficult.) Within large domains of a fixed preferred spin orientation, electrons are
completely delocalized, so that the ground state energy is {\it not} separated from the
energies of excited states by a uniformly positive energy gap.

\medskip
\subsection*{Acknowledgments} We are grateful to a referee for very useful comments and
suggestions.

\section{Setting, models, and summary of main results}

We consider a lattice model with electrons in two different bands, indexed by $a=1,2$. We
let $c_{a\sigma}^\dagger(x)$ and $c_{a\sigma}(x)$ denote the creation and annihilation operators
for an electron of band $a$ and spin $\sigma=\uparrow,\downarrow$, at site
$x\in\bbZ^d$. The state space of the system in a finite domain $\Lambda \subset \bbZ^d$ is
the Hilbert space
\be
\caH_\Lambda = \caF_\Lambda \otimes \caF_\Lambda,
\ee
where $\caF_\Lambda$ is the usual Fock space for electrons,
\be
\caF_\Lambda = \bigoplus_{N\geq0} P_- \Bigl[ \ell^2(\Lambda) \otimes \bbC^2 \Bigr]^{\otimes
N}.
\ee
Here $P_-$ is the projector onto antisymmetric functions. The energy of electrons is
partly kinetic and partly due to interactions among themselves. The kinetic energy is
represented by standard hopping terms. Interactions are of two different kinds. First,
Coulomb interactions are approximated by on-site operators of the Hubbard type. Second, a
pair interaction involving the spins of electrons of different bands reminds of the Hund
rule for the filling of atomic eigenstates. Precisely, we define the `2-band Hund-Hubbard
Hamiltonian' as
\be
\begin{split}
\label{defHam}
H_\Lambda^{\rm HH} = &-\sum_{a=1,2} t_a \sum_{\sigma=\uparrow,\downarrow} \sumtwo{x,y \in
\Lambda}{|x-y|=1} c_{a\sigma}^\dagger(x) c_{a\sigma}(y) + \sum_{a=1,2} U_a
\sum_{x\in\Lambda} n_{a\uparrow}(x) n_{a\downarrow}(x) \\
&+ U_{12} \sum_{x\in\Lambda} n_1(x) \, n_2(x) - J \sum_{x\in\Lambda} \bsS_1(x) \cdot \bsS_2(x).
\end{split}
\ee
The parameters $t_a$ control the kinetic energy of electrons of each band and they are
related to the effective mass of electrons. We suppose that $t_1>t_2$ and refer to
electrons of band 1 as `light' (they move fast) and electrons of band 2 as `heavy'
(they move slowly). The parameters $U_1,U_2,U_{12}$ are positive and represent the Coulomb
interaction
energy between two particles of band 1, two particles of band 2, and one particle of each
band, respectively. The number operators are defined by $n_{a\sigma}(x) = c_{a\sigma}^\dagger(x)
c_{a\sigma}(x)$ and $n_a(x) = n_{a\uparrow}(x) + n_{a\downarrow}(x)$. Finally, $J>0$
measures the strength of the coupling between the spins; spin operators
are given by
\be
S_a^{(j)}(x) = \sum_{\sigma,\sigma'} \tau_{\sigma\sigma'}^{(j)} c_{a\sigma}^\dagger(x)
c_{a\sigma'}(x),
\ee
where $\tau^{(j)}$, $j=1,2,3$, are the Pauli matrices $\frac12 (\begin{smallmatrix}
0&1\\1&0 \end{smallmatrix})$, $\frac12 (\begin{smallmatrix} 0&-\ii\\\ii&0
\end{smallmatrix})$, $\frac12 (\begin{smallmatrix} 1&0\\0&-1 \end{smallmatrix})$.
We will consider a simpler model with anisotropic
spin interactions (the third components of the spins interact
more strongly) and with the property that one of the two bands
is nearly flat. The Hamiltonian of this model is introduced in
\eqref{defaHHM}, below. The following discussion applies to general
Hund-Hubbard models, isotropic and anisotropic ones.

We note that neither the hopping terms alone nor the Hund couplings alone give
rise to global magnetization. Without on-site interactions the ground state favored by
the hopping terms is not magnetic; the kinetic energy is
minimized by a state where half the electrons have spin $\uparrow$, and half the
electrons have
spin $\downarrow$. As for the Hund couplings, they are local (on-site) and do not yield
the long-range correlations that are present in a ferromagnetic state. Ferromagnetism in
this model results from a cooperation of the two terms.

This model was numerically studied in \cite{HV1} for $t_1=t_2$. Ferromagnetic phases were
identified at low temperatures and for some intervals of electronic densities. The present
model with $t_1<t_2$ actually behaves more like the two-band Hubbard model with Kondo
spins of \cite{HV2}; heavy electrons here play a r\^ole similar to that of Kondo spins.
Ferromagnetic phases were also numerically observed for this model.

In this paper, we study a phase with spontaneous magnetization. The total spin operator
in a domain $\Lambda\subset\bbZ^d$ is denoted $\bsM_\Lambda$ and is given by
\be
\bsM_\Lambda = \sum_{x\in\Lambda} \bigl[ \bsS_1(x) + \bsS_2(x) \bigr].
\ee
The norm of $\bsM_\Lambda$ is
\be
\|\bsM_\Lambda\| = \Bigl( \sum_{i=1}^3 (M_\Lambda^{(i)})^2 \Bigr)^{1/2},
\ee
with $M_\Lambda^{(i)}$ the $i$-th component of $\bsM_\Lambda$. We expect that the
system displays extensive magnetization. That is, in a domain $\Lambda$ the expectation of $\|\bsM_\Lambda\|$ in the ground
state should be proportional to $|\Lambda|$. We are not able to prove this, but we can
prove that the system is magnetized at least on a `mesoscopic' scale. To be precise, we
consider the following definition of the magnetization per site:
Given a state $\Upsilon \in \caH_\Lambda$ and a subdomain $\Lambda' \subset \Lambda$, we define
\be
\label{defm}
\begin{split}
m_{\Lambda,\Lambda'} &= \frac1{|\Lambda|} \sum_{x: \Lambda'+x
\subset \Lambda} \frac1{|\Lambda'|} \bigl( \Upsilon, \| \bsM_{\Lambda'+x} \| \Upsilon
\bigr); \\
m_{\Lambda,\Lambda'}^{(3)} &= \frac1{|\Lambda|} \sum_{x: \Lambda'+x
\subset \Lambda} \frac1{|\Lambda'|} \bigl| (\Upsilon, M_{\Lambda'+x}^{(3)} \Upsilon)
\bigr|.
\end{split}
\ee
Note that we consider the expectation value of the norm of $\bsM_{\Lambda'}$ in the state
$\Upsilon$, averaged over all translates of $\Lambda'$ such that $\Lambda'+x$ remains in
$\Lambda$. The number of terms in the sum is $|\Lambda|$, up to a negligible boundary
correction. We clearly have that $m_{\Lambda,\Lambda'} \geq m_{\Lambda,\Lambda'}^{(3)}$. We
will prove that, for the anisotropic model introduced in \eqref{defaHHM}, below, the third component of the ground state magnetization
per site, $m_{\Lambda,\Lambda'}^{(3)}$, satisfies
$$
\lim_{\Lambda' \nearrow \bbZ^d} \lim_{\Lambda \nearrow \bbZ^d} m_{\Lambda,\Lambda'}^{(3)} > 0.
$$
The limits are over boxes of increasing size, and electron densities are kept constant.

Perturbation methods shed some light onto the structure of the phases of this model. The
situation is similar to the one in
the Hubbard model at half-filling and with strong on-site repulsion, which behaves like the
antiferromagnetic Heisenberg model.
Perturbative methods along the lines of \cite{DFFR,DFF} can be used for many rational densities.
However, such methods can be applied
only when electrons are localized, and this is not the case in a conducting
metal. Here we focus our attention on regimes where there does not exist an energy gap
separating excitations from the
ground state energy, and where some electrons have delocalized wave functions.

The Hamiltonian \eqref{defHam} is symmetric with respect to spin rotations, and this makes
the study difficult. We simplify the model by replacing the term $\bsS_1(x) \cdot
\bsS_2(x)$ by $S_1^{(3)}(x) \cdot S_2^{(3)}(x)$. Furthermore, we let $t_2\to0$.
We can fix the energy scale by choosing $t_1=1$. We then obtain the following simplified
Hamiltonian (`2-band Ising-Hubbard model')
\be
\begin{split}
H_\Lambda^{\rm IH} = &-\sum_\sigma \sumtwo{x,y \in
\Lambda}{|x-y|=1} c_{1\sigma}^\dagger(x) c_{1\sigma}(y) + \sum_{a=1,2} U_a \sum_{x\in\Lambda}
n_{a\uparrow}(x) n_{a\downarrow}(x) + U_{12} \sum_{x\in\Lambda} n_1(x) n_2(x) \\
&- \frac J4 \sum_{x\in\Lambda} [n_{1\uparrow}(x) - n_{1\downarrow}(x)] \cdot
[n_{2\uparrow}(x) - n_{2\downarrow}(x)].
\label{defIHM}
\end{split}
\ee

Let $\rho_1$, $\rho_2$ denote the densities of light and heavy particles, respectively.
We prove the following statement in Section \ref{secIHM}. Our proof works in
dimension larger or equal to 2 (Equation \eqref{2donly} holds for $d\geq2$ only).
We do not have results for the one-dimensional models.

\begin{theorem}
\label{thmIHM}
Let $d\geq2$.
For arbitrary $0<\rho_1<\rho_2\leq1$, there exists $J_0<\infty$ such that, for $\frac J4 -
U_{12} > J_0$, all ground states $\Upsilon$ of $H_\Lambda^{\rm IH}$ satisfy
$$
\lim_{\Lambda' \nearrow \bbZ^d} \lim_{\Lambda \nearrow \bbZ^d} m_{\Lambda,\Lambda'}^{(3)} =
\frac{\rho_1+\rho_2}2.
$$
\end{theorem}

This theorem suggests that the ground state displays `saturated
ferromagnetism', as it has maximum total spin. Notice that for large enough $J$ it
holds independently of $U_1, U_2 \geq 0$.

The proof of Theorem \ref{thmIHM} reduces to the study of the ground state energy for fixed configurations of heavy
electrons, since the latter do not have kinetic energy. To a configuration of heavy electrons
we can assign domains of $\uparrow$ and $\downarrow$ spins. A light electron of spin
$\uparrow$, say, is in a state that is essentially localized on the
domain where heavy electrons also have spin $\uparrow$. Hund interactions suppress other
configurations. In the limit $J\to\infty$, the ground state energy is purely kinetic and it is
minimal in a configuration of heavy electrons with large domains of identical spins. This
allows us
to show that the size of the boundary of these domains is less than $|\Lambda|^{1-\frac1d}$, meaning
that spins are locally aligned. See Section \ref{secIHM} for details.

The constant $J_0$ in Theorem \ref{thmIHM} depends on $\rho_1$, and there are good
reasons for it. There is no ferromagnetism for $\rho_1=0$ or $\rho_1=\rho_2=1$. The former case
results in independent spins at each site. The latter case can be
treated with perturbation methods. Non-empty sites are typically occupied by two particles of
spin $\uparrow$, or by two particles of spin $\downarrow$. An effective
interaction of strength $\frac2{U_1 + \frac J2}$ stabilizes antiferromagnetic chessboard phases in the
ground state and at low temperatures. This interaction can be obtained using the method
described in \cite{DFFR}. The case $\rho_1=\rho_2<1$ is more subtle.

We can improve the result of Theorem \ref{thmIHM} and consider a model that
interpolates between \eqref{defHam} and \eqref{defIHM}. We refer to the following
Hamiltonian as `the asymmetric 2-band Hund-Hubbard model':
\be
\label{defaHHM}
\begin{split}
H_\Lambda^{\rm aHH} = &-\sum_{a=1,2} t_a \sum_{\sigma=\uparrow,\downarrow} \sumtwo{x,y \in
\Lambda}{|x-y|=1} c_{a\sigma}^\dagger(x) c_{a\sigma}(y) + \sum_{a=1,2} U_a
\sum_{x\in\Lambda} n_{a\uparrow}(x) n_{a\downarrow}(x) \\
&+ U_{12} \sum_{x\in\Lambda} n_1(x) \, n_2(x) - J \sum_{x\in\Lambda} S_1^{(3)}(x)
S_2^{(3)}(x) \\
&- J^\perp \sum_{x\in\Lambda} \bigl( S_1^{(1)}(x)S_2^{(1)}(x) +
S_1^{(2)}(x)S_2^{(2)}(x) \bigr).
\end{split}
\ee

Let us again fix the energy scale by setting $t_1=1$.

\begin{theorem}
\label{thmaHHM}
Let $d\geq2$.
For arbitrary $0<\rho_1<\rho_2\leq1$, there are constants $J_0<\infty$ and $c>0$ (both
depend on the densities) such that if $\frac J4 - U_{12} > J_0$ and $t_2,J^\perp
< c$, all ground states $\Upsilon$ of $H_\Lambda^{\rm aHH}$ satisfy
$$
\lim_{\Lambda' \nearrow \bbZ^d} \lim_{\Lambda \nearrow \bbZ^d} m_{\Lambda,\Lambda'} =
\frac{\rho_1+\rho_2}2.
$$
\end{theorem}

This theorem again holds uniformly in $U_1, U_2 \geq 0$.

Our paper is organized as follows. We discuss the properties of the ground state of a
simple model in Section \ref{secbounds}. The results for the simple model are then used in
Section \ref{secIHM} where Theorem \ref{thmIHM} is proved. Finally, it is shown in Section
\ref{secaHHM} that the claims for the Ising-Hubbard Hamiltonian \eqref{defIHM} can be
extended to certain perturbations, that include the asymmetric Hund-Hubbard model
\eqref{defaHHM}. This proves Theorem \ref{thmaHHM}.

\section{Interacting electrons in a magnetic potential}
\label{secbounds}

We introduce in this section a Hubbard model of electrons in an external potential that
involves the third components of the spins. We do not insist on the physical relevance of
this model. The sole motivation for this section stems from applications to Hund-Hubbard
systems. We will use Propositions \ref{propbounds1}--\ref{propbounds2} in Sections
\ref{secIHM} and \ref{secaHHM} in order to prove Theorems \ref{thmIHM} and \ref{thmaHHM} ---
these theorems being physically motivated.

The results below extend the bounds for the ground state energy of spinless electrons in
binary potentials proposed in \cite{FLU}. We work in the Fock space $\caF_\Lambda$ of spin
$\frac12$ electrons in $\Lambda$. Let $V$ be a `magnetic potential', that is,
$V$ is a collection of nonnegative numbers $V_x^\uparrow, V_x^\downarrow$ indexed by sites
$x\in\bbZ^d$. The Hamiltonian is
\be
\label{defHmagnetic}
H_\Lambda(V) = -\sumtwo{x,y\in\Lambda}{|x-y|=1} \sum_{\sigma=\uparrow,\downarrow}
c_\sigma^\dagger(x) c_\sigma(y) + U \sum_{x\in\Lambda} n_\uparrow(x) n_\downarrow(x) +
\sum_{x\in\Lambda} \sum_{\sigma=\uparrow,\downarrow} V_x^\sigma \, n_\sigma(x).
\ee
Here, $c_\sigma^\dagger(x)$ and $c_\sigma(x)$ are creation and annihilation operators of
fermions of spin $\sigma$ at $x$, and $n_\sigma(x) = c_\sigma^\dagger(x) c_\sigma(x)$. We
suppose that a gap $V_0$ separates the minimum value from other values of the potential.
Introducing
\be
\label{defA}
A_\sigma = \{ x\in\Lambda: V_x^\sigma=0 \}, \quad A = A_\uparrow \cup A_\downarrow
\ee
(the sites where the potential is zero for some spin), we define
\be
V_0 = \min_{\sigma=\uparrow,\downarrow} \inf_{x\notin A_\sigma} V_x^\sigma.
\ee
We assume that $V_0$ is strictly positive.

In order to understand the bounds on the ground state energy given below, it is useful to
consider the situation where $V_0\to\infty$. Assuming that $A_\uparrow \cap A_\downarrow =
\emptyset$, the domain $\Lambda$ is partitioned into $A_\uparrow$, $A_\downarrow$, and
$\Lambda\setminus A$. Electrons of spin $\sigma$ are described by wave functions with
support in $A_\sigma$, the
energy being infinite otherwise. Electrons do not interact and their ground state energy
is purely kinetic. It mainly consists of a bulk term that depends on the electronic
density inside $A_\sigma$ and that is proportional to the volume $|A_\sigma|$. The effect of
the boundary of $A_\sigma$ is to increase the ground state energy by a term proportional
to the size of the
boundary. The ground state energy of non-interacting spinless electrons in arbitrary
finite domains was studied in \cite{FLU}; upper and lower bounds were established that
confirm the discussion above.
As $V_0$ decreases from infinity to a finite value, electrons delocalize somewhat, but the situation
does not change in any essential way.

Estimates for the ground state energy involve the energy density of free spinless
electrons in the limit of infinite volume. As is well-known, the energy per site $e(\rho)$
for a density $0<\rho<1$ of electrons is given by
\be
e(\rho) = \frac1{(2\pi)^d} \int_{\varepsilon_k < \varepsilon_{\rm F}(\rho)} \varepsilon_k
\, \dd k, \quad\quad \varepsilon_k = -2\sum_{i=1}^d \cos k_i,
\ee
where $\varepsilon_{\rm F}(\rho)$ is the Fermi energy, defined by the equation
\be
\rho = \frac1{(2\pi)^d} \int_{\varepsilon_k < \varepsilon_{\rm F}(\rho)} \dd k.
\ee
Notice that $e(\rho)<0$ for $0<\rho<1$. We need to define the boundary $B(A)$ of a set $A \subset
\bbZ^d$; it is convenient to define it as the number of bonds that connect $A$ with its
complement,
\be
B(A) = \# \{ (x,y): x\in A, y\notin A, |x-y|=1 \}.
\ee

We first give a bound for fixed densities of electrons of each spin. In the absence of
interactions ($U=0$) the following proposition merely rephrases similar results in
\cite{FLU}. We define $E_\Lambda(V;N_\uparrow,N_\downarrow)$ as the ground state energy of
$H_\Lambda(V)$ when the number of spin $\uparrow$ (spin $\downarrow$) electrons is
$N_\uparrow$ ($N_\downarrow$ respectively). We introduce a notation for electronic densities
inside $A_\uparrow$ and $A_\downarrow$; for $\sigma = \uparrow,\downarrow$, we let $\rho_\sigma =
\frac{N_\sigma}{|A_\sigma|}$. We have the following bounds for the ground state energy.

\begin{proposition}
\label{propbounds1}
Let $V$ be a magnetic potential and $N_\uparrow$, $N_\downarrow$ be numbers with the
properties that:
\begin{itemize}
\item $A_\uparrow \cap A_\downarrow = \emptyset$.
\item $V_0 > 2d(\sqrt d+1)$.
\item $N_\uparrow \leq |A_\uparrow|$, $N_\downarrow \leq |A_\downarrow|$.
\end{itemize}
Then there exists $\alpha(\rho)>0$ (independent of $V_0$), for $0 < \rho <
\frac{|A|}{|\Lambda|}$, such that
\ba
\sum_{\sigma=\uparrow,\downarrow} \Bigl[ e(\rho_\sigma) |A_\sigma| -
\tfrac{e(\rho_\sigma)}{2d} B(A_\sigma) \Bigr] &\geq E_\Lambda(V;N_\uparrow,N_\downarrow)
\nn \\
&\geq
\sum_{\sigma=\uparrow,\downarrow} \Bigl[ e(\rho_\sigma) |A_\sigma| + \bigl(
\alpha(\rho_\sigma) - \gamma(V_0) \bigr) B(A_\sigma) \Bigr] \nn
\end{align}
with $\gamma(V_0) = \tfrac{4d}{V_0-2d} + \tfrac{16d^3}{(V_0-2d)^2 - 4d^3}$.
\end{proposition}

The inequalities in this proposition hold uniformly in $U$.
The proof of Proposition \eqref{propbounds1} is based on results in ref.\
\cite{FLU}, where the sum, $S_{\Lambda,N}$, of the $N$ lowest eigenvalues of
the discrete Laplacian $t_{xy} = -\delta_{|x-y|,1}, x,y\in\Lambda$ is estimated, with $\Lambda$ a finite
set of lattice points of arbitrary shape. Two of the results in \cite{FLU}
are relevant for our analysis:
\begin{itemize}
\item We have upper and lower bounds,
\be
\label{boundsFLU}
e(\rho) |\Lambda| - \tfrac{e(\rho)}{2d} B(\Lambda) \geq S_{\Lambda,N} \geq e(\rho)
|\Lambda| + a(\rho) B(\Lambda),
\ee
where $\rho = \frac N{|\Lambda|}$, and $a(\rho)$ is strictly positive for any $0<\rho<1$.
Recall that $e(\rho)$ is negative, so that all boundary terms in the above equation are
positive.
(The notation in \cite{FLU} is slightly different, the Hamiltonian being shifted by $2d$
and the boundary is defined differently.)
\item If $S_{\Lambda,N}^U$ denotes the sum of the $N$ lowest eigenvalues of the operator
$-\delta_{|x-y|,1} + U \upchi_{\Lambda^\compl}(x)$, where $\upchi_{\Lambda^\compl}$ is the
characteristic function of the complement, $\Lambda^\compl$, of the set $\Lambda$, and if $U$ is positive, we have
\be
\label{bound2FLU}
S_{\Lambda,N} \geq S_{\Lambda,N}^U \geq S_{\Lambda,N} - \gamma(U) B(\Lambda),
\ee
for some $\gamma(U) \to 0$ as $U\to\infty$.
\end{itemize}

The upper bound for $E_\Lambda(V;N_\uparrow,N_\downarrow)$ does not depend on
$V_0$. Increasing the values of the potential actually incresases the energy, so it is
enough to prove the statement in the limit $V_0\to\infty$. Electrons with different
spins are independent, and the upper bound follows from the one in \eqref{boundsFLU}.

Let us turn to the lower bound.
The operator that represents interactions between electrons is positive; we get a lower
bound for the ground state energy by taking $U\to0$. For $V_0=\infty$ we are in the situation of
\cite{FLU}. For finite $V_0$ we use \eqref{bound2FLU} with minor modifications. Namely,
starting with Equations (4.3)--(4.5) of \cite{FLU} but introducing our measure
$B(\Lambda)$ of the boundary, the upper bound in Eq.\ (4.8) can be replaced by
$\frac{4d}{V_0-2d} B(\Lambda)$.
It is useful to modify the bound for the number of sites at distance $n$ from the domain
$\Lambda$ (recall that we are using the $\ell^1$ distance here). It is not hard to
check that
\be
\begin{split}
&\# \{ x: \dist(x,\Lambda) = 1 \} \leq B(\Lambda);\\
&\# \{ x: \dist(x,\Lambda) = n \} \leq d \cdot \# \{ x: \dist(x,\Lambda) = n-1 \} \quad
\text{if } n\geq2.
\end{split}
\ee
We therefore have that $\# \{ x: \dist(x,\Lambda) = n \} \leq d^{n-1} B(\Lambda)$. This allows
to bound $N - \Tr\tilde\rho$ in Eq.\ (4.11) by $\frac{4d^2}{(V_0-2d)^2 - 4d^3} B(\Lambda)$,
leading to the present definition of $\gamma(V_0)$. The bound given here is better for
large $V_0$ than the one in \cite{FLU}.

The considerations above show that Proposition \ref{propbounds1} is a mild extension of
\cite{FLU}. The following proposition needs, however, a more detailed proof.

\begin{proposition}
\label{propoverlap}
Under the same hypotheses as in Proposition \ref{propbounds1},
we have that, for all normalized ground states $\Upsilon$ of $H_\Lambda(V)$,
$$
\sum_{\sigma=\uparrow,\downarrow} \sum_{x\notin A_\sigma} (\Upsilon, n_\sigma(x) \Upsilon)
\leq \tfrac3{V_0-4d} [B(A_\uparrow) + B(A_\downarrow)].
$$
\end{proposition}

\begin{proof}
Let $M_\sigma$ be the number of electrons of spin $\sigma$ that are outside of
$A_\sigma$, and $P_{M_\uparrow M_\downarrow}$ be the projector onto the subspace
spanned by states with exactly $M_\sigma$ particles outside $A_\sigma$,
$\sigma=\uparrow,\downarrow$. A state $\Upsilon$ can be decomposed as
\be
\label{decomposition}
\Upsilon = \sum_{M_\uparrow=0}^{N_\uparrow} \sum_{M_\downarrow=0}^{N_\downarrow} c_{M_\uparrow
M_\downarrow} \Upsilon_{M_\uparrow M_\downarrow},
\ee
with $c_{M_\uparrow M_\downarrow} = \| P_{M_\uparrow M_\downarrow} \Upsilon \| \geq
0$, $\sum c^2_{M_\uparrow M_\downarrow} = 1$, and $\Upsilon_{M_\uparrow M_\downarrow} =
c_{M_\uparrow M_\downarrow}^{-1} P_{M_\uparrow M_\downarrow} \Upsilon$ is normalized. The goal is to estimate
\be
\label{expectation}
\sum_{\sigma=\uparrow,\downarrow} \sum_{x\notin A_\sigma} (\Upsilon, n_\sigma(x) \Upsilon)
= \sum_{M_\uparrow, M_\downarrow} c^2_{M_\uparrow M_\downarrow} (M_\uparrow + M_\downarrow).
\ee
The strategy is to obtain a lower bound for $E_\Lambda(V;N_\uparrow,N_\downarrow)$ that
involves the expression above. Comparison with the upper bound of Proposition
\ref{propbounds1} will prove the claim.

The ground state energy is increasing in $U$ so that we can again set $U=0$ when
discussing a lower bound. The Hamiltonian $H_\Lambda(V)$ can be split into
\be
\label{splitHam}
H_\Lambda(V) = \sum_{\sigma=\uparrow,\downarrow} \Bigl( H_{A_\sigma}^\sigma(V) +
H_{\Lambda\setminus A_\sigma}^\sigma(V) \Bigr) - \sum_{\sigma=\uparrow,\downarrow} \sumtwo{x\in A_\sigma, y\notin A_\sigma}{|x-y|=1}
\bigl[ c_\sigma^\dagger(x) c_\sigma(y) +
c_\sigma^\dagger(y) c_\sigma(x) \bigr].
\ee
Hamiltonians $H_{\boldsymbol\cdot}^\sigma(V)$ consist in kinetic terms for particles of spin $\sigma$
in the corresponding domains, and of the potentials given by $V^\sigma$. They leave
the subspace with fixed $M_\uparrow$ and $M_\downarrow$ invariant.
The norm of the last operator is smaller than $2B(A_\uparrow) + 2B(A_\downarrow)$.
Therefore
\bm
\label{voila1}
(\Upsilon, H_\Lambda(V) \Upsilon) \geq \sum_{M_\uparrow,M_\downarrow} c^2_{M_\uparrow
M_\downarrow} \sum_\sigma (\Upsilon_{M_\uparrow M_\downarrow}, [H_{A_\sigma}^\sigma(V) +
H_{\Lambda \setminus A_\sigma}^\sigma(V)] \Upsilon_{M_\uparrow M_\downarrow}) \\
- 2B(A_\uparrow) - 2B(A_\downarrow).
\end{multline}
Inserting the lower bound for the sum of the lowest eigenvalues of the
discrete Laplacian in a finite domain, neglecting the positive boundary
correction term, we get the lower bound
\be
\label{voila2}
(\Upsilon_{M_\uparrow M_\downarrow}, H_{A_\sigma}^\sigma(V) \Upsilon_{M_\uparrow
M_\downarrow}) \geq e(\tfrac{N_\sigma-M_\sigma}{|A_\sigma|}) |A_\sigma| \geq e(\rho_\sigma)
|A_\sigma| - \varepsilon_{\rm F}(\rho_\sigma) M_\sigma.
\ee
The second inequality holds because $e(\rho+\eta) \geq e(\rho) + \eta \varepsilon_{\rm
F}(\rho)$ (indeed, $e(\rho)$ is convex and its derivative is $\varepsilon_{\rm F}(\rho)$).

The Hamiltonian $H_{\Lambda \setminus A_\sigma}^\sigma(V)$ is the second-quantized version
of a one-body Hamiltonian, whose eigenvalues are bigger than $V_0-2d$. Since
$\varepsilon_{\rm F}(\rho) \leq 2d$, we have
\be
\label{voila3}
(\Upsilon, H_\Lambda(V) \Upsilon) \geq \sum_{\sigma=\uparrow,\downarrow} e(\rho_\sigma)
|A_\sigma| + (V_0-4d) \sum_{M_\uparrow,M_\downarrow} c^2_{M_\uparrow M_\downarrow}
(M_\uparrow + M_\downarrow) - 2B(A_\uparrow) - 2B(A_\downarrow).
\ee
The right side must be less than the upper bound for
$E_\Lambda(V;N_\uparrow,N_\downarrow)$ stated in Proposition \ref{propbounds1}. Using
$-\frac{e(\rho)}{2d} \leq 1$, we get Proposition \ref{propoverlap}.
\end{proof}

We turn to the situation where the total number of electrons is specified, but not their
spins. Let $E_\Lambda(V;N)$ be the ground state energy of $H_\Lambda(V)$ with $N$ electrons.
In the proof of the following proposition we have to assume
that the dimension of the system is at least 2.

\begin{proposition}
\label{propbounds2}
We suppose $d\geq2$.
Let $V$ be a magnetic potential and $N$ be a number, that satisfy
\begin{itemize}
\item $A_\uparrow \cap A_\downarrow = \emptyset$.
\item $V_0 > 2d(\sqrt d+1)$.
\item $N \leq |A|$.
\end{itemize}
Let $\rho = \frac N{|A|} < 1$; then there exists $\bar\alpha(\rho)>0$ such that
$$
e(\rho) |A| - \tfrac{e(\rho)}{2d} [B(A_\uparrow) + B(A_\downarrow)] \geq E_\Lambda(V;N) \geq
e(\rho) |A| + \bigl( \bar\alpha(\rho) - \gamma(V_0) \bigr) [B(A_\uparrow) +
B(A_\downarrow)].
$$
\end{proposition}

\begin{proof}
The upper bound follows from the upper bound of Proposition \ref{propbounds1} that holds
for all $U$. We can set $U=0$ for the lower bound. Because
electrons of different spins do not interact, the ground state energy is given by a sum of
lowest eigenvalues of the corresponding one-body Hamiltonians for particles of given spin.
Let $N_\uparrow$ be the number of spin $\uparrow$ electrons in the ground state. Taking
into account multiplicities, there are $|A_\uparrow|$
available eigenvalues in $(-2d,2d)$ for spin $\uparrow$ electrons, and $|A_\downarrow|$
eigenvalues for spin $\downarrow$ electrons. Other eigenvalues are larger than $V_0-2d$.
Since $N \leq |A|$, we must have $0 \leq N_\uparrow \leq |A_\uparrow|$ and $0 \leq
N - N_\uparrow \leq |A_\downarrow|$.

Let us introduce $\rho = \frac N{|A|}$, $\rho' = \frac{N_\uparrow}{|A|}$, and $\eta
= \frac{|A_\uparrow|}{|A|}$. We have $\rho_\uparrow = \frac{\rho'}\eta$ and
$\rho_\downarrow = \frac{\rho-\rho'}{1-\eta}$. Using the lower bound of Proposition
\ref{propbounds1}, we obtain
\bm
E_\Lambda(V;N) \geq \bigl\{ \eta e(\tfrac{\rho'}\eta) + (1-\eta)
e(\tfrac{\rho-\rho'}{1-\eta}) \bigr\} |A| + \bigl[
\alpha(\tfrac{\rho'}\eta) - \gamma(V_0) \bigr] B(A_\uparrow) \\
+ \bigl[
\alpha(\tfrac{\rho-\rho'}{1-\eta}) - \gamma(V_0) \bigr] B(A_\downarrow).
\end{multline}
This bound does not hold for all $\rho'$, but it holds when $\rho'$ corresponds to a
ground state. We get a lower bound by minimizing over $\rho'$. A difficulty arises, namely
that the coefficient of $B(A_\uparrow)$ or of $B(A_\downarrow)$ could be negative. The term in braces reaches
its minimum for $\frac{\rho'}\eta = \frac{\rho-\rho'}{1-\eta} = \rho$. Let $\epsilon$
be such that the minimizer for the whole right side be $\frac{\rho'}\eta = \rho -
\frac\epsilon\eta$ (and $\frac{\rho-\rho'}{1-\eta} = \rho + \frac\epsilon{1-\eta}$).
The fractions $\frac\epsilon\eta$ and $\frac\epsilon{1-\eta}$ are small, because $\alpha$
and $\gamma$ are small. Hence $\epsilon$ is small, too.
Let $f(\epsilon)$ denote the term in braces,
\be
f(\epsilon) = \eta e(\rho - \tfrac\epsilon\eta) + (1-\eta) e(\rho +
\tfrac\epsilon{1-\eta}).
\ee
The second derivative is
\be
f''(\epsilon) = \tfrac1\eta \varepsilon_{\rm F}'(\rho - \tfrac\epsilon\eta) +
\tfrac1{1-\eta} \varepsilon_{\rm F}'(\rho + \tfrac\epsilon{1-\eta}).
\ee
One easily verifies that $\varepsilon_{\rm F}'(\rho) \geq c$, with $c$ strictly
positive when $d\geq2$. This implies that
\be
\label{2donly}
f(\epsilon) \geq e(\rho) + \tfrac c{2\eta(1-\eta)} \epsilon^2.
\ee
Then
\bm
E_\Lambda(V;N) \geq e(\rho) |A| + \tfrac{c\epsilon^2}{\eta(1-\eta)} |A|
+ \bigl[ \alpha(\rho - \tfrac\epsilon\eta) - \gamma(V_0) \bigr]
B(A_\uparrow) \\
+ \bigl[ \alpha(\rho + \tfrac\epsilon{1-\eta}) - \gamma(V_0) \bigr]
B(A_\downarrow).
\end{multline}
The right side should be $e(\rho) |A| + \bar\alpha [B(A_\uparrow) + B(A_\downarrow)]$. We
must show that the brackets are strictly positive, depending on $\rho$ and $V_0$, but uniformly in
$\eta$. Four situations need to be carefully investigated: (1) if
$\rho$ is small and $\frac\epsilon\eta>0$ is of the order of
$\rho$; (2) if $\rho$ is small and $\frac\epsilon{1-\eta}<0$ is of the order of
$\rho$; (3) if $\rho$ is close to 1 and $\frac\epsilon\eta<0$ is of the order of
$1-\rho$; and (4) if $\rho$ is close to 1 and $\frac\epsilon{1-\eta}>0$ is of the order of
$\rho$. These four cases are similar, so it is enough to consider case (1). The factor
in front of $B(A_\downarrow)$ is bounded away from 0 because $\rho +
\frac\epsilon{1-\eta}$ is bounded away from 0 and 1 (uniformly in $\eta$), so that
$\alpha(\rho + \frac\epsilon{1-\eta}) > 0$ (uniformly in $\eta$). We can assume that
$\frac\epsilon\eta > \frac\rho2$ (the bound is uniform in $\eta$ otherwise), and we
consider the factor in front of $B(A_\uparrow)$. We take advantage of the second term in
$|A|$, observing that
\be
\tfrac{c\epsilon^2}{\eta(1-\eta)} |A| > \tfrac c{1-\eta}
(\tfrac\rho2)^2 \eta|A|.
\ee
Now $\eta|A| = |A_\uparrow| \geq \frac1{2d} B(A_\uparrow)$, and we see that the factor in front of $B(A_\uparrow)$ is uniformly bounded away
from zero as $\eta\to0$.
\end{proof}

\section{The Ising approximation}
\label{secIHM}

In this section we prove Theorem \ref{thmIHM}. Heavy electrons are static and
they can be treated classically. Their state is represented by a classical spin
configuration $s_\Lambda \in \{ 0, \uparrow, \downarrow, 2 \}^\Lambda$, and the model
\eqref{defIHM} corresponds to a Hamiltonian, $H_\Lambda(s_\Lambda)$, acting on
$\caF_\Lambda$. The expression for $H_\Lambda(s_\Lambda)$ is given by
\eqref{defIHM}, with the understanding that the operators $c_{1\sigma}^\dagger(x),
c_{1\sigma}(x)$ act on $\caF_\Lambda$ (instead of $\caF_\Lambda \otimes \caF_\Lambda$),
and the operators $n_{2\sigma}(x)$ are replaced by numbers as follows:
\ba
&n_{2\uparrow}(x) \mapsto \begin{cases} 1 &\text{if } s_x=\uparrow \\ 0 &\text{if }
s_x=\downarrow, \end{cases} \nn\\
&n_{2\downarrow}(x) \mapsto \begin{cases} 0 &\text{if
} s_x=\uparrow \\ 1 &\text{if } s_x=\downarrow. \end{cases} \nn
\end{align}
Thus $H_\Lambda(s_\Lambda)$ is a Hubbard Hamiltonian with an external potential (or
`field') given by $s_\Lambda$. It is convenient to add a constant $\frac J4 - U_{12}$ to
the energy so that the potential is non-negative. For given $s_\Lambda$ we define the
potential $V^\uparrow$ by
\be
\label{defpotential}
V^\uparrow_x = \begin{cases} 0 & \text{if } s_x = \uparrow \\ \frac J2 & \text{if } s_x =
\downarrow \\ \frac J4 - U_{12} & \text{if } s_x = 0 \\ \frac J4 + U_{12} & \text{if } s_x
= 2. \end{cases}
\ee
Next, we define $V^\downarrow$ in the same way, by flipping the spins.
With $N_1 = \rho_1 |\Lambda|$, the Hamiltonian for the Ising-Hubbard model can be
expressed using the Hamiltonian $H_\Lambda(V)$ defined in \eqref{defHmagnetic},
namely
\be
\label{newdefIHM}
H_\Lambda(s_\Lambda) + (\tfrac J4 - U_{12}) N_1 = H_\Lambda(V)
+ U_2 \sum_{x\in\Lambda} \delta_{s_x,2}.
\ee

The strategy of our proof of Theorem \ref{thmIHM} is as follows:
\begin{itemize}
\item A state where all electrons have spin $\uparrow$ gives us an upper bound for the
ground state energy (Equation \eqref{bornesup}).
\item We derive a lower bound for the ground state energy that involves `classical
excitations' of $s_\Lambda$ --- regions where heavy particles do not have parallel spins. See
Proposition \ref{propIHM}.
\item By combining the upper and lower bounds for the ground state energy, we find that
any ground state configuration necessarily has only few excitations (see Eq.\ \eqref{bornebord}). This
suffices to prove Theorem \ref{thmIHM}.
\end{itemize}

Let $E(s_\Lambda;\rho_1)$ denote the ground state energy of \eqref{newdefIHM}, and let $N_2$
be the number of heavy electrons in $s_\Lambda$.
A candidate for the ground state is a purely ferromagnetic state, where all particles have
spin $\uparrow$. Heavy electrons occupy a domain $A_\uparrow$ with $|A_\uparrow|=N_2$, and
they are described by the configuration $s_x=\uparrow$ for all
$x\in A_\uparrow$. Light electrons also have spin $\uparrow$; electrons of identical spins do
not interact, so that
the ground state is given by the $\rho_1 |\Lambda|$ lowest eigenstates of the hopping
matrix in $\Lambda$. By the upper bound in Proposition \ref{propbounds1}, we have that
\be
\min_{s_\Lambda} 
E(s_\Lambda;\rho_1) \leq e(\tfrac{\rho_1}{\rho_2}) N_2 + 4d N_2^{1-\frac1d}.
\label{bornesup}
\ee
The ratio $\frac{\rho_1}{\rho_2} = \frac{N_1}{N_2}$ represents the effective
density of light electrons when they all reside in $A_\uparrow$.
The second term on the right side is an upper bound for the boundary contribution to the
energy of an optimal domain
with $N_2$ sites.

Next, we turn to a lower bound. It is useful to introduce
\be
\label{defxi}
\xi(\rho) = \rho \varepsilon_{\rm F}(\rho) - e(\rho).
\ee
Notice that $0\leq\xi(\rho)\leq2d$, and $\xi'(\rho) = \rho \varepsilon_{\rm F}'(\rho)>0$ so
that $\xi(\rho)$ is increasing. Recall the definition \eqref{defA} for sets $A_\uparrow$
and $A_\downarrow$ that are determined by the potential \eqref{defpotential}. Notice that $A$
is the set of sites occupied by exactly one heavy particle, and that
$A_\uparrow \cap A_\downarrow = \emptyset$.

\begin{proposition}
\label{propIHM}
For $0<\rho_1<\rho_2\leq1$, there exist $J_0<\infty$ and $\widetilde\alpha>0$ (both independent
of $\Lambda$) such that if $\frac J4 - U_{12} >
J_0$, we have that
$$
E(s_\Lambda;\rho_1) \geq  e(\tfrac{\rho_1}{\rho_2}) N_2 + \bigl[ \xi(\tfrac{\rho_1}{\rho_2}) +
\tfrac12 U_2 \bigr] (N_2-|A|) + \widetilde\alpha \bigl[ B(A_\uparrow) + B(A_\downarrow)
\bigr],
$$
for arbitrary $s_\Lambda$.
\end{proposition}

{\it Remark:} A similar bound can be proven when $0<\rho_1<2-\rho_2\leq1$.

One main consequence of Proposition \ref{propIHM} can be obtained by combining it
with the upper bound \eqref{bornesup}. We get
\be
\label{bornebord}
\begin{cases} |A| \geq N_2 (1 - \const \cdot N_2^{-1/d}) \\ B(A_\uparrow) + B(A_\downarrow) \leq
\const \cdot N_2^{1-\frac1d} \end{cases}
\ee
for constants that are uniform in the size of the system. These inequalities imply Theorem
\ref{thmIHM}, as is shown below.

\begin{proof}[Proof of Proposition \ref{propIHM}]
We observe that $\sum_x \delta_{s_x,2} = \frac12(N_2-|A|)$ yielding the term involving $U_2$.
To alleviate our notation we suppose now that $U_2=0$.
Let $\zeta>0$ be a small number; we first consider configurations such that
$\frac{N_1}{|A|} \leq 1-\zeta$. Proposition \ref{propbounds2} gives
\be
\label{encoreuneborne}
E(s_\Lambda;\rho_1) \geq e(\tfrac{N_1}{|A|}) |A| + \bigl( \bar\alpha(\tfrac{N_1}{|A|}) -
\gamma(\tfrac J4 - U_{12}) \bigr) [B(A_\uparrow)+B(A_\downarrow)].
\ee
The function $\nu e(\frac\rho\nu)$ is convex in $\nu$, and its derivative with respect to
$\nu$ is equal to $-\xi(\frac\rho\nu)$. Therefore
\be
e(\tfrac{N_1}{|A|}) |A| = e(\tfrac{\rho_1}{\rho_2} \tfrac{N_2}{|A|}) \tfrac{|A|}{N_2} N_2 \geq
e(\tfrac{\rho_1}{\rho_2}) N_2 + \xi(\tfrac{\rho_1}{\rho_2}) (N_2-|A|).
\ee

Since $\frac{N_1}{|A|} < 1-\zeta$ the function
$\bar\alpha(\tfrac{N_1}{|A|})$ is uniformly bounded away from zero, and we obtain a
strictly positive $\widetilde\alpha$, provided $\frac J4 - U_{12}$ is large enough (see
Proposition \ref{propbounds1}).

We now consider configurations such that $1-\zeta < \frac{N_1}{|A|} \leq 1$. Equation
\eqref{encoreuneborne} is still valid but $\bar\alpha(\frac{N_1}{|A|})$ may be very small
and we ignore it; it is positive. Convexity of $e(\rho)$ yields
\be
e(\tfrac{N_1}{|A|}) \geq e(1) + (\tfrac{N_1}{|A|}-1) \varepsilon_{\rm F}(1) \geq -2d\zeta.
\ee
We have used that $e(1)=0$ and $\varepsilon_{\rm F}(1)=2d$. Because $B(A_\uparrow)+B(A_\downarrow)
\leq 2d|A|$, we obtain from \eqref{encoreuneborne}
\be
E(s_\Lambda;\rho_1) \geq -2d \bigl[ \zeta + \gamma(\tfrac J4 - U_{12}) + \widetilde\alpha
\bigr] |A| + \widetilde\alpha \bigl[ B(A_\uparrow) + B(A_\downarrow) \bigr].
\ee
In order to complete the proof of Proposition \ref{propIHM}, we need to check that
\be
-2d \bigl[ \zeta + \gamma(\tfrac J4 - U_{12}) + \widetilde\alpha \bigr] |A| \geq
e(\tfrac{\rho_1}{\rho_2}) N_2 + \xi(\tfrac{\rho_1}{\rho_2}) (N_2-|A|).
\ee
We have that $N_2-|A| = N_2 [1 - \frac{\rho_1}{\rho_2} \frac{|A|}{N_1}] = N_2 [1 -
\frac{\rho_1}{\rho_2} + O(\zeta)]$. As $|A| \leq N_2$ and because the term in brackets can
be arbitrary small (depending on $\rho_1$, $\rho_2$), it is enough to check that
\be
0 > e(\tfrac{\rho_1}{\rho_2}) + \xi(\tfrac{\rho_1}{\rho_2}) (1 - \tfrac{\rho_1}{\rho_2}).
\ee
Using the definition \eqref{defxi} of $\xi$, the condition can be reduced to $\xi(\rho) -
\varepsilon_{\rm F}(\rho) > 0$ for $0<\rho<1$. This is easy to verify, as this function is
strictly decreasing and $\xi(1)-\varepsilon_{\rm F}(1)=0$.

Finally, the case where $N_1>|A|$ is easy because we can use Proposition \ref{propIHM} for
the lowest $|A|$ eigenvalues, and remaining eigenvalues are larger than $\frac J4 - U_{12}
- 2d \geq 0$.
\end{proof}

\begin{proof}[Proof of Theorem \ref{thmIHM}]
We have established inequalities \eqref{bornebord} that show that ground state
configurations of heavy electrons consist of large domains with one particle of spin
$\uparrow$ at each site, large domains with one particle of spin $\downarrow$, or domains void
of particles. Boundaries of these domains are `sparse'. Recall that definition
\eqref{defm} of the magnetization $m_{\Lambda,\Lambda'}$ involves an average over
translates of $\Lambda'$. It is enough to restrict to boxes that are fully in $A_\uparrow$
or in $A_\downarrow$. Indeed, few boxes are intersecting their boundaries, and there are
virtually no electrons outside of $A$.

It is clear that $m_{\Lambda,\Lambda'}^{(3)} \leq \frac{\rho_1+\rho_2}2$ for all states with
densities $\rho_1$ and $\rho_2$ of light and heavy electrons, so that it suffices to
establish the converse inequality. The definition of $m_{\Lambda,\Lambda'}^{(3)}$ involves
a sum over translates of $\Lambda'$ that are inside $\Lambda$. All terms are positive, so we get a lower
bound by restricting the sum to translates that are contained in either $A_\uparrow$ or
$A_\downarrow$:
\be
\sum_{x: \Lambda'+x \subset \Lambda} \bigl| (\Upsilon, M_{\Lambda'+x}^{(3)} \Upsilon)
\bigr| \geq
\sum_{x: \Lambda'+x \subset A_\uparrow} (\Upsilon, M_{\Lambda'+x}^{(3)} \Upsilon) -
\sum_{x: \Lambda'+x \subset A_\downarrow} (\Upsilon, M_{\Lambda'+x}^{(3)} \Upsilon).
\label{restrictedsum}
\ee
Let us recall the definition of $M_{\Lambda'}^{(3)}$:
\be
M_{\Lambda'}^{(3)} = \sum_{x\in\Lambda'} M^{(3)}(x),
\ee
with
\be
\label{ilfautnumeroterca}
M^{(3)}(x) = \tfrac12 \sum_{a=1,2} \bigl( n_{a\uparrow}(x) - n_{a\downarrow}(x) \bigr).
\ee
Since $\| M^{(3)}(x) \| \leq 1$, we have from \eqref{restrictedsum}
\be
\label{caaussi}
\sum_{x: \Lambda'+x \subset \Lambda} (\Upsilon, |M_{\Lambda'+x}^{(3)}| \Upsilon) \geq
\sum_{x \in A_\uparrow} (\Upsilon, M^{(3)}(x) \Upsilon) - \sum_{x \in A_\downarrow}
(\Upsilon, M^{(3)}(x) \Upsilon)  - |\Lambda'| [B(A_\uparrow) + B(A_\downarrow)].
\ee

Let $N_{1\uparrow} = \sum_{x\in\Lambda} (\Upsilon, n_{1\uparrow}(x) \Upsilon)$; then
\be
\label{etcaaussi}
\sum_{x \in A_\uparrow} (\Upsilon, M^{(3)}(x) \Upsilon) = \tfrac12 |A_\uparrow| + \tfrac12
N_{1\uparrow} - \sum_{x\notin A_\uparrow} (\Upsilon, n_{1\uparrow}(x) \Upsilon).
\ee
The latter term is less than $\frac3{\frac J4 - U_{12} - 4d} B(A_\uparrow)$ by Proposition
\ref{propoverlap}. The same argument applies to spin $\downarrow$ electrons. Using
inequalities \eqref{bornebord}, we see that $m_{\Lambda,\Lambda'}$ is larger than $\frac12
\rho_2 + \frac12 \rho_1$, up to a term of order $|\Lambda|^{-1/d}$ (it depends on
$\Lambda'$). This term vanishes in the limit $\Lambda \nearrow \bbZ^d$.
\end{proof}

\section{The asymmetric Hund-Hubbard model}
\label{secaHHM}

We now turn to the proof of Theorem \ref{thmaHHM}.
The asymmetric Hund-Hubbard model \eqref{defaHHM} can be expressed as a perturbation of
the Ising-Hubbard model \eqref{defIHM}. Namely, with $t_1=1$,
\be
H_\Lambda^{\rm aHH} = H_\Lambda^{\rm IH} - t_2 \sum_{\sigma=\uparrow,\downarrow}
\sumtwo{x,y\in\Lambda}{|x-y|=1} c_{2\sigma}^\dagger(x) c_{2\sigma}(y)
- J^\perp \sum_{x\in\Lambda} \bigl[ S_1^{(1)}(x)
S_2^{(1)}(x) + S_1^{(2)}(x) S_2^{(2)}(x) \bigr].
\ee
In the previous section we showed that any ground state configuration of the Ising-Hubbard
model satisfies inequalities \eqref{bornebord}. Heavy electrons are now quantum particles and a
classical configuration cannot be an eigenstate. We can extend \eqref{bornebord} by
expanding the ground state in the basis of configurations of heavy particles, and show
that \eqref{bornebord} holds in average. Namely, we denote by $\Phi(s_\Lambda) \in
\caF_\Lambda$ the normalized state of heavy electrons in the configuration $s_\Lambda$. Clearly,
$(\Phi(s_\Lambda))$ is a basis of $\caF_\Lambda$. Any state $\Upsilon \in \caF_\Lambda
\otimes \caF_\Lambda$ has a unique decomposition as
\be
\label{stateexpansion}
\Upsilon = \sum_{s_\Lambda} c(s_\Lambda) \Psi(s_\Lambda) \otimes \Phi(s_\Lambda),
\ee
where $c(s_\Lambda) \geq 0$ satisfies $\sum_{s_\Lambda} c^2(s_\Lambda) = 1$, and
$\Psi(s_\Lambda)$ is {\it some} normalized state that represents the light particles.
Notice the asymmetry in notation: $\Psi(s_\Lambda)$ is indexed by $s_\Lambda$, but the
configuration of spins of light particles may be very
different from the configuration $s_\Lambda$, in general. In particular, $\Psi(s_\Lambda)$
describes a state with $N_1$ particles, while $\Phi(s_\Lambda)$ has $N_2$ particles.
Let $X(s_\Lambda)$ denote the number of `excitations' of $s_\Lambda$; namely,
\be
X(s_\Lambda) = (N_2 - |A|) + B(A_\uparrow) + B(A_\downarrow).
\ee
The extension of \eqref{bornebord} is as follows.

\begin{proposition}
\label{propaHHM}
Let $d\geq2$, and $0<\rho_1<\rho_2\leq1$. There are constants $J_0<\infty$ and $\gamma>0$
such that if $\frac J4 - U_{12} > J_0$ and $t_2,J^\perp
< \gamma$, and if $c(s_\Lambda)$ are the
coefficients defined in \eqref{stateexpansion} for a ground state of $H_\Lambda^{\rm aHH}$, we
have
$$
\sum_{s_\Lambda} c^2(s_\Lambda) X(s_\Lambda) \leq \const \cdot N_2^{1-\frac1d},
$$
for a constant that is independent of $\Lambda$.
\end{proposition}

\begin{proof}
We again define $V$ by \eqref{defpotential} and
$H_\Lambda(V)$ by \eqref{newdefIHM}. $U_2$ plays no r\^ole here, just as in Section
\ref{secIHM}; so we set it to 0 from now on.
By the variational principle we find an upper bound for the ground state energy by
considering a state where all electrons have spin $\uparrow$. Heavy electrons are
packed together and light ones are in appropriate delocalized wave functions with support
on $A_\uparrow$. Neither the $t_2$ term nor the $J^\perp$ term contributes to the energy of
this state, and \eqref{bornesup} therefore continues to be an upper bound for the ground state
energy.

The goal is now to find a lower bound with the same bulk term as in the equation above,
plus a correction that involves the average of $X(s_\Lambda)$. For $\Upsilon$ expanded as in
\eqref{stateexpansion}, we have
\be
\label{valeurmoyenne}
\begin{split}
(\Upsilon, H_\Lambda^{\rm aHH} \Upsilon) &+ (\tfrac J4 - U_{12}) N_1 = \sum_{s_\Lambda} c^2(s_\Lambda) \Bigl( \Psi(s_\Lambda),
H_\Lambda(V) \Psi(s_\Lambda) \Bigr) \\
&- t_2 \sum_{s_\Lambda, s_\Lambda'} \sum_{\sigma = \uparrow,\downarrow}
\sumtwo{x,y\in\Lambda}{|x-y|=1} c(s_\Lambda) c(s_\Lambda') \Bigl( \Psi(s_\Lambda),
\Psi(s_\Lambda') \Bigr) \Bigl( \Phi(s_\Lambda), c_{2\sigma}^\dagger(x) c_{2\sigma}(y)
\Phi(s_\Lambda') \Bigr) \\
&- J^\perp \sum_{s_\Lambda, s_\Lambda'} \sum_{x\in\Lambda} c(s_\Lambda) c(s_\Lambda')
\Bigl( \Psi(s_\Lambda) \otimes \Phi(s_\Lambda), \\
&\hspace{35mm} \bigl[ S_1^{(1)}(x) S_2^{(1)}(x) +
S_1^{(2)}(x) S_2^{(2)}(x) \bigr] \Psi(s_\Lambda') \otimes \Phi(s_\Lambda') \Bigr).
\end{split}
\ee
The first term involves the same $H_\Lambda(V)$ that appears in \eqref{newdefIHM}; this
gives us the right bulk contribution. The other two terms are actually
irrelevant and it is enough to find estimates.

We observe that the second term on the right side of \eqref{valeurmoyenne} is less than
\be
\label{boundkinheavy}
t_2 \sum_{s_\Lambda, s_\Lambda'}{}' c(s_\Lambda) c(s_\Lambda') \leq t_2 \Bigl(
\sum_{s_\Lambda, s_\Lambda'}{}' c^2(s_\Lambda) \Bigr)^{1/2} \Bigl(
\sum_{s_\Lambda, s_\Lambda'}{}' c^2(s_\Lambda') \Bigr)^{1/2} = t_2
\sum_{s_\Lambda,s_\Lambda'}{}' c^2(s_\Lambda).
\ee
Primed sums are over configurations that are identical except for a heavy
electron moved to a neighboring site. Given $s_\Lambda$, there are less than $2d(N_2-|A|) +
B(A_\uparrow) + B(A_\downarrow)$ such configurations $s_\Lambda'$ (recall that $\frac12
(N_2-|A|)$ is the number of sites that are occupied by two heavy electrons). It follows
that \eqref{boundkinheavy} is smaller than $2d t_2 \sum_{s_\Lambda}
c^2(s_\Lambda) X(s_\Lambda)$ (and it is larger than the negative of this expression).

The third term of the right side of \eqref{valeurmoyenne} can be treated in the same spirit. It is necessary
to cast the perpendicular Hund interactions in a form that shows that their contribution
is no more than the boundary between domains of identical spins. We
therefore introduce standard operators $S_a^{(+)}(x)$, $S_a^{(-)}(x)$, by
\be
\begin{split}
& S_a^{(+)}(x) = S_a^{(1)}(x) + \ii S_a^{(2)}(x), \\
& S_a^{(-)}(x) = S_a^{(1)}(x) - \ii S_a^{(2)}(x).
\end{split}
\ee
Perpendicular spin interactions become
\be
S_1^{(1)}(x) S_2^{(1)}(x) + S_1^{(2)}(x) S_2^{(2)}(x) = \tfrac12 \bigl[ S_1^{(+)}(x)
S_2^{(-)}(x) + S_1^{(-)}(x) S_2^{(+)}(x) \bigr].
\ee

Let $x\in A$, and $s_\Lambda^x$ be the configuration obtained from $s_\Lambda$ by flipping
the spin at $x$. The third term of \eqref{valeurmoyenne} is equal to
\be
\label{troisieme}
-\tfrac12 J^\perp \sum_{s_\Lambda} \sum_{x\in A} c(s_\Lambda) c(s_\Lambda^x) \Bigl(
\Psi(s_\Lambda), S_1^{(\#)}(x) \Psi(s_\Lambda^x) \Bigr),
\ee
with $\#=$`$+$' if $s_x=\downarrow$, and $\#=$`$-$' if $s_x=\uparrow$. Since $S_1^{(+)}(x) =
c_{1\uparrow}^\dagger(x) c_{1\downarrow}(x)$, the Schwarz inequality yields the bound
\be
\Bigl| \Bigl( \Psi(s_\Lambda), S_1^{(+)}(x) \Psi(s_\Lambda^x) \Bigr) \Bigr| \leq \Bigl(
\Psi(s_\Lambda), n_{1\uparrow}(x) \Psi(s_\Lambda) \Bigr)^{1/2} \Bigl(
\Psi(s_\Lambda^x), n_{1\downarrow}(x) \Psi(s_\Lambda^x) \Bigr)^{1/2}.
\ee
A similar inequality holds when $S_1^{(+)}(x)$ is replaced with $S_1^{(-)}(x)$; one should
simply interchange $n_{1\uparrow}(x)$ and $n_{1\downarrow}(x)$ on the right side. Then
echoing \eqref{boundkinheavy}, the absolute value of \eqref{troisieme} is found to be smaller than
\bm
\label{boundJterm}
\tfrac{J^\perp}2 \Bigl[ \sum_{s_\Lambda} \sum_{x\in A} c^2(s_\Lambda) \Bigl(
\Psi(s_\Lambda), n_{1,-s_x}(x) \Psi(s_\Lambda) \Bigr) \Bigr]^{1/2} \\
{\boldsymbol\cdot} \Bigl[ \sum_{s_\Lambda}
\sum_{x\in A} c^2(s_\Lambda^x) \Bigl( \Psi(s_\Lambda^x), n_{1,s_x}(x) \Psi(s_\Lambda^x)
\Bigr) \Bigr]^{1/2} \\
= \tfrac{J^\perp}2 \sum_{s_\Lambda} \sum_{x\in A} c^2(s_\Lambda) \Bigl(
\Psi(s_\Lambda), n_{1,-s_x}(x) \Psi(s_\Lambda) \Bigr).
\end{multline}
This expression is reminiscent of the expression in Proposition \ref{propoverlap}. However, $\Psi(s_\Lambda)$ is
{\it not} a ground state for light electrons in the magnetic potential given by
$s_\Lambda$ and therefore the proposition does not directly apply.

We can nevertheless recycle the ideas underlying the proof of Proposition \ref{propoverlap}.

Let $\caN = (N_\uparrow, N_\downarrow, M_\uparrow, M_\downarrow)$ be four positive
integers such that $N_\uparrow + N_\downarrow = N_1$, and $M_\sigma \leq N_\sigma$.
$N_\sigma$ is the number of light electrons of spin $\sigma$, and $M_\sigma$ is the number
of light electrons of spin $\sigma$ that are not localized on the favorable sites
$A_\sigma$. We can expand $\Psi(s_\Lambda)$ according to $\caN$, in a fashion that is
reminiscent of \eqref{decomposition},
\be
c(s_\Lambda) \Psi(s_\Lambda) = \sum_\caN c_\caN(s_\Lambda) \Psi_\caN(s_\Lambda)
\ee
where coefficients are positive and states are normalized. With this notation we observe
that \eqref{boundJterm} is bounded above by the following expression
similar to \eqref{expectation}
\be
\label{boundJperp}
\tfrac{J^\perp}2 \sum_{s_\Lambda} \sum_\caN c^2_\caN(s_\Lambda) (M_\uparrow +
M_\downarrow).
\ee
The Hamiltonian $H_\Lambda(V)$ can be split as in \eqref{splitHam} and we obtain the lower
bound \eqref{voila1} with $c(s_\Lambda) \Psi(s_\Lambda)$ in lieu of $\Upsilon$, and
$c_\caN(s_\Lambda) \Psi_\caN(s_\Lambda)$ in lieu of $c_{M_\uparrow M_\downarrow}
\Upsilon_{M_\uparrow M_\downarrow}$. We then get \eqref{voila2} and \eqref{voila3}.
Explicitly, the lower bound for \eqref{valeurmoyenne} is
$$
e(\tfrac{\rho_1}{\rho_2}) N_2 + \sum_{s_\Lambda} \sum_\caN c_\caN^2(s_\Lambda) \Bigl\{ (\tfrac J4 - U_{12} - 4d -
\tfrac{J^\perp}2) (M_\uparrow + M_\downarrow) - 2B(A_\uparrow) - 2 B(A_\downarrow) - 2d
t_2 X(s_\Lambda) \Bigr\}.
$$
The bulk term $e(\frac{\rho_1}{\rho_2}) N_2$ above comes from \eqref{voila3}, minimizing
over $N_\uparrow$ and $N_\downarrow$. This expression is less than the upper bound
\eqref{bornesup}; this implies that
\be
\label{cettefoiscestlabonne}
\sum_{s_\Lambda} \sum_\caN c_\caN^2(s_\Lambda) (M_\uparrow + M_\downarrow) \leq \frac{4d +
2 + 2d t_2}{\frac J4 - U_{12} - 4d - \frac{J^\perp}2} \sum_{s_\Lambda} c^2(s_\Lambda)
X(s_\Lambda).
\ee
(We used that $N_2^{1-\frac1d} \leq X(s_\Lambda)$.)
Notice that the factor on the right side is small. This
estimate is necessary to bound \eqref{boundJperp}.

Using Proposition \ref{propIHM} for the first term in \eqref{valeurmoyenne}, we then
conclude that
\be
\sum_{s_\Lambda} c^2(s_\Lambda) \Bigl( \Psi(s_\Lambda), H_\Lambda(V)
\Psi(s_\Lambda) \Bigr) \geq e(\tfrac{\rho_1}{\rho_2}) N_2 + \widetilde\alpha \sum_{s_\Lambda}
c^2(s_\Lambda) X(s_\Lambda).
\ee
(We assume here that $\widetilde\alpha$ is smaller than $\xi(\frac{\rho_1}{\rho_2}) + \frac12 U_2$.)
Again invoking the upper bound \eqref{bornesup}, we obtain
\be
e(\tfrac{\rho_1}{\rho_2}) N_2 + 4d N_2^{1-\frac1d} \geq e(\tfrac{\rho_1}{\rho_2}) N_2 +
\Bigr[ \widetilde\alpha - 2d t_2 - \frac{J^\perp}2 \, \frac{4d +
2 + 2d t_2}{\frac J4 - U_{12} - 4d - \frac{J^\perp}2} \Bigr] \sum_{s_\Lambda} c^2(s_\Lambda)
X(s_\Lambda).
\ee
The quantity in brackets is strictly positive when $\frac J4 - U_{12}$ is large enough,
and this proves Proposition \ref{propaHHM}.
\end{proof}

\begin{proof}[Proof of Theorem \ref{thmaHHM}]
The proof is similar to that of Theorem \ref{thmIHM}, except that we use Proposition
\ref{propaHHM} instead of the inequalities \eqref{bornebord}. All equations until
\eqref{ilfautnumeroterca} hold without change. Eqs \eqref{caaussi} and \eqref{etcaaussi}
need to be modified because the $A_\sigma$'s are not fixed here. These equations hold when
averaged over $s_\Lambda$ with weights $c^2(s_\Lambda)$. We obtain
\ba
\sum_{x, \Lambda'+x \subset \Lambda} (\Upsilon, |M_{\Lambda'+x}^{(3)}| \Upsilon) &\geq
\sum_{s_\Lambda} \sum_\caN c^2_\caN(s_\Lambda) \sum_\sigma \Bigl[ \tfrac12 |A_\sigma| +
\tfrac12 N_\sigma - \tfrac12 M_\sigma - |\Lambda'| B(A_\sigma) \Bigr] \nn\\
&\geq \tfrac12 (N_1+N_2) - 2 |\Lambda'| \sum_{s_\Lambda} c^2(s_\Lambda) X(s_\Lambda).
\end{align}
We have used \eqref{cettefoiscestlabonne} in order to estimate the contribution of
$M_\uparrow$ and $M_\downarrow$. After division by $|\Lambda|$, the last term vanishes in
the thermodynamic limit by Proposition \ref{propaHHM}.
\end{proof}

\end{document}